# The Network of Mexican Cities

R. Mendozas[1]    R. Mansilla[1,2,*]

[1] Centro de Investigaciones Interdisciplinarias en Ciencias y Humanidades, Torre II de Humanidades, 4to piso, Ciudad Universitaria, México 04510 D.F.
[2] Centro de Ciencias de la Complejidad, Torre de Ingeniería, 5to piso, Ciudad Universitaria México 04510 D.F.

## Abstract

The network of 5823 cities of Mexico with a population more than 5000 inhabitants is studied. Our analysis is focused to the spectral properties of the adjacency matrix, the small-world properties of the network, the distribution of the clustering coefficients and the degree distribution of the vertices. The connection of these features with the spread of epidemics on this network is also discussed.

[*] Whom correspondence should be addressed: mansy@servidor.unam.mx

**I Introduction**

In the last decade we have witnessed a growing interest in complex networks [1]-[6]. This type of web like structures, allow us the study of a wide variety of systems, ranging from networks of chemicals connected by chemical reactions to the Internet and the World Wide Web. Among these structures, very often cited examples come from social science and specifically from studies of the distribution of human settlements [7]. This last subject plays an important roll in the analysis of the spread of epidemics in human populations.

The aim of this paper is the study of the most important features of the network of Mexican cities with a population equal or greater than 5000 inhabitants. They amount to 5823 cities. Two cities are considered connected if there is a road, highway, internal flight or coastal shipping connecting them *directly*, meaning without any intermediate city among them. The network was built using data drawn from the INEGI Database [8].

The paper is organized as follows. In Section II the results related to the adjacency matrix are shown. Some findings on the spectra of this matrix are discussed. In Section III and IV the small world properties of the network are established. In Section III the features of shortest path lengths between nodes are studied. In Section IV the distribution of the clustering coefficients are analyzed. In Section V the properties of degree distribution of the vertices are discussed. Section VI is for discussion of the relationship of the results obtained and the spread of epidemics on Mexico and other concluding remarks.

**II The Adjacency Matrix**.

In mathematical terms a network can be represented by a graph $G = \{P, E\}$, where $P$ is the set of vertices or nodes (in our case the cities) and $E$ is the set of edges or links (in our case the collection of roads, highways, internal flights or coastal transport lines connecting two cities directly). To simplify the structure, in our network not more than one edge may connect a given pair of nodes. When there are different forms to connect two cities, we take only one edge.

A useful representation of the network is the adjacency matrix $M$. The dimension of this matrix is $N \times N$, where $N = 5823$ is the number of nodes of the network. Entries $M_{ij}$ are binary numbers defined according the rule:

$$M_{ij} = \begin{cases} 1 & \text{if node } i \text{ is connected with node } j \\ 0 & \text{otherwise} \end{cases}$$

Obviously this matrix is symmetric with diagonal elements $M_{ii} = 0$.

In the INEGI Database every city has a code number. These code numbers are allocated on the basis of two criteria. The first is the alphabetical order of the state to which the city belongs. The second one is the geographical proximity among the cities. This code univocally characterizes every city. These numbers are used to order the rows and columns of the entries in matrix $M$. The use of this codification has the advantage that cities geographically close have in general close entries in the adjacency matrix.

The Fig. 1 shows a representation of the matrix $M$. The dots mean non-zero entries in the matrix. Labels are inserted for some states of the federation. The horizontal line labeled with number 9 represents the direct connections of Mexico City to other cities, which amount to 65 links. Because of the matrix $M$ is symmetric, the corresponding vertical line also exists. This line is related to the dominant eigenvector which is strongly localized on this hub (see bellow in this Section). Compare this figure with Fig. 1 of [9].

The matrix $M$ has $N$ eigenvalues $\lambda_i$. The spectral density is defined as follows:

$$\rho(\lambda) = \frac{1}{N} \sum_{i=1}^{N} \delta(\lambda - \lambda_i)$$

Obviously, as $N \to +\infty$ the spectral density $\rho(\lambda)$ tends to a continuous function. The study of properties of this spectrum has received a wide attention in the literature. It is well known that some topological features of networks are close related with the spectral density [10], [11]. Let us consider, for instance, a random graph $G_{N,p}$ with a connection probability:

$$p(N) = \frac{A}{N^{\alpha}} \quad , \quad \alpha < 1$$

As $N \to +\infty$ the spectral density approaches the continuous function:

$$\rho(\lambda) = \begin{cases} \frac{\sqrt{4Np(1-p) - \lambda^2}}{2\pi Np(1-p)} & , \quad |\lambda| < 2\sqrt{Np(1-p)} \\ 0 & , \quad \text{otherwise} \end{cases}$$

Which is known as Wigner's law or semicircle law (see [10], pp. 75, 78, 93, 135, 166, 274, 371, 400, 420). A graph of the above function can be seen in Fig. 1 of [12].

The spectral density of the adjacency matrix is shown in Fig. 2a. The horizontal axis of the figure is rescaled as $\lambda/\sqrt{Np(1-p)}$ while the vertical axis is rescaled as $\rho\sqrt{Np(1-p)}$ where the parameter $p$ is the "implied probability of connection" defined as follows: In Section V it is shown that the standard deviation of degree centrality $k$ is $\sigma_k = 3.3963$. On the other hand $\sigma_k = \sqrt{Np(1-p)}$. Using the last relation $p = 0.0020$ is found. According to [12] the largest eigenvalue $\lambda_1$ satisfies $\lim_{N \to +\infty} \frac{\lambda_1}{N} = p$. In our case $\lambda_1 = 9.728610283$ which yields the result $\lambda_1/N = 9.728610283/5823 = 0.0017$ also in good agreement with the value obtained above.

In Fig. 2b a comparison of the spectral density and semicircle law is shown. The scale in the vertical axis is logarithmic in order to enhance the features of the densities. Notice the difference between Fig. 2b of this paper and Fig. 3 of [12], where spectral densities of small world networks and semicircle law are shown. It is interesting to remark that no skewness is observed in the spectral density. However, in Section III it is shown that our network exhibit properties characteristic of small world networks.

A more sophisticated measure of the centrality of nodes is the eigenvector centrality [13]-[14]. This measure indicates that not all connections are equally important. The connections to highly connected nodes are more important than those with nodes having few connections. If we denote the eigenvector centrality of the node $i$ by $x_i$, then the vector $\mathbf{x} = (x_1, \ldots, x_N)$ satisifes $M\mathbf{x} = \lambda_1 \mathbf{x}$, where $\lambda_1$ is the largest eigenvalue of the matrix $M$. According to the Perron-Frobenius theorem, all components of $\mathbf{x}$ are nonnegative.

In Fig. 3 the components of the eigenvector $\mathbf{x}$ are shown. Unlike [14] we simply plot the components of the dominant eigenvector instead of the squared components of it. Labels are inserted for some cities. Notice the strong localization of this vector in the component corresponding to Mexico City. The importance of this property is discussed further.

**III Small world properties**.

Two limiting-case network topologies have been extensively reported in the literature: ordered and random networks. But it is well known that empirical evidence suggests many social, technological and biological networks lie somewhere in-between these two extremes. In this Section we show that our network has properties of small world topology, i.e., every two nodes can be connected by going through only a short chain of intermediate nodes. In the following Section the clustering properties of our network are studied: two nodes with a common connection are far more likely to be also connected. Both features are the fingerprints of the small world topology, (see, for instance, Definition 4.1.3 of [1] or [15]-[16]).

Table I shows the distribution of the shortest paths. From the total possible number of pairs of cities $N(N-1)/2 = 16950753$ there are 16649481 paths. Therefore 52 little towns are disconnected from the giant component of the network. The matrix $M$ has only 10312 entries different from zero, hence this is the number of shortest path of length one. The diameter of the network is 27.

Let us denote by $l$ the random variable shortest path length. Then, using the data of the Table I, its expected value is $\langle l \rangle = 8.7146$. On the other hand $\ln N = \ln 5823 = 8.6696$. Both values differ in less than 0.52%. It is well known [5], [9], [17]-[19] that $\langle l \rangle \approx \ln N$ is a distinguishing feature of the small world topology.

In Fig. 4a the probability density function of shortest path lengths is shown. Notice that there is almost no skewness in the distribution (recall $\langle l \rangle = 8.7146$). As stated in Section II the eigenvector centrality is strongly localized in the hub corresponding to Mexico City. It induces to assume that the model proposed in [20] should fit well to our network. In Fig. 4b the probability density function of shortest path lengths is shown as well as the best fit of the empirical data to the theoretical distribution (Equation (13) of [20]) proposed in that paper. It implies that the influence of other hubs, although much less stronger than Mexico City's, plays a decisive role on the behavior of the distribution of shortest path length.

**IV The clustering coefficients.**

In their seminal work [17] Watts and Strogatz defined the small world behavior on the basis of two ingredients. The first is the property that two nodes can be connected by going through only a short chain of intermediate nodes. The second one has a local character and is referred as the "cliquishness" among the nodes. If node A is connected with nodes B and C then is highly probable that also B and C are connected.

We define first the clustering coefficient for nodes (see for instances [3], [21], [22]): if $k_i$ is the degree of node $i$ then there are $k_i(k_i-1)/2$ possible connections among its neighbors. Let's $e_i$ be the actual number of connections. The clustering coefficient $C_i$ for node $i$ is defined as:

$$C_i = \frac{2e_i}{k_i(k_i-1)}$$

The clustering coefficient of the network is defined as the average of clustering coefficients of the nodes $C = \langle C_i \rangle$.

In Fig. 5a, the probability density function of the clustering coefficients is shown. Notice that about the 4% of the nodes are cliques, i.e., all its neighbors are connected among them. The 3% of the nodes have clustering coefficient with a value about $1/3$.

Fig. 5b shows the average value of clustering coefficients as a function of the degree of the corresponding nodes. Notice that there are several values of the degree $k$ which have average clustering coefficient zero. It implies a tree like behavior for these nodes.

It is well known that in random networks the clustering coefficient $C_{rand} \approx \langle k \rangle / N$ [4], [16]. In contrast, in our network $\langle k \rangle / N = 0.000613$ and also $C = \langle C_i \rangle = 0.068$. Therefore the clustering coefficient of the network is two orders of magnitude greater than the corresponding value of a random network. It is also a well known property of small world topology.

**V The degree distribution of nodes.**

The probability density function of node degrees is one of the most studied properties of a network. The interest on this trait resides mainly in the fact that real networks such as the Internet have what is called scale free distribution [23], [24]. On the other hand, random graphs studied by Erdös and Rényi [25] have Poissonian degree distribution in the limit of large graph size. It is also known that the degree distribution of small world model does not match the behavior observed in real world [5].

In Fig. 6a is shown the probability density function of the node degrees. The values corresponding to Mexico City and Durango were removed because they look as outliers. No scale free or other known behavior is observed. As suggested in [5] the cumulative probability function was also calculated:

$$P(k > z) = \sum_{k>z} p_k$$

where $p_k$ is the probability of observe a node with degree $k$ on the network. A good fit to a Beta-like function [26], [27] were found for the cumulative distribution however:

$$P(k > z) = 0.0323 \frac{(24-z)^{1.4551}}{z^{1.513}}$$

The identification of coefficients was done by linear regression on the logarithmic variables. The fitness of the model was $R^2 = 0.9545$. In Fig. 6b is shown the empirical and the theoretical cumulative distributions.

The expected value of degrees is $\langle k \rangle = 3.5732$ and the standard deviation $\sigma_k = 3.3963$. See Section II for the relevance of this last number.

**VI Concluding remarks.**

The small world structure of the network of Mexican cities with population equal or greater than 5000 inhabitants is established. To our knowledge, no previous work of this type has been reported (although not updated see Table II of [5] or Chapter III of [4]). This work was originally motivated by the possibility of the spreading of infectious diseases (such as AH1N1). Obviously, transport is one of the most common functions of networked systems. Therefore some concluding remarks on this topic could be done.

In a remarkable paper [28] the hidden metric structure of networks was established and the general mechanism of navigability was identified, based on the concept of similarity between nodes. Using the clustering coefficient of the network and the scaling exponent of the degree distribution the authors define a boundary between the navigable region and the non navigable region (see Fig. 3d of [28]). Although we could not identify a power law in the degree distribution and hence neither a scaling exponent, the clustering coefficient obtained in Section IV seems to suggest that the network of cities is placed in the non navigable region. Recently we were made aware of the paper [29] where the authors study the dynamics of information spreading not detached to the process of epidemic spreading.

The authors would like to thanks to P. Miramontes and G. Cocho for some fruitful discussions, to N. Del Castillo for provide important advices in the programming tasks, to J. L. Gordillo of DGSCA, the supercomputing facility of UNAM for his support. Last but not least, the author would like to thank E. Ruelas of the Secretary of Health for his long standing support of this project.

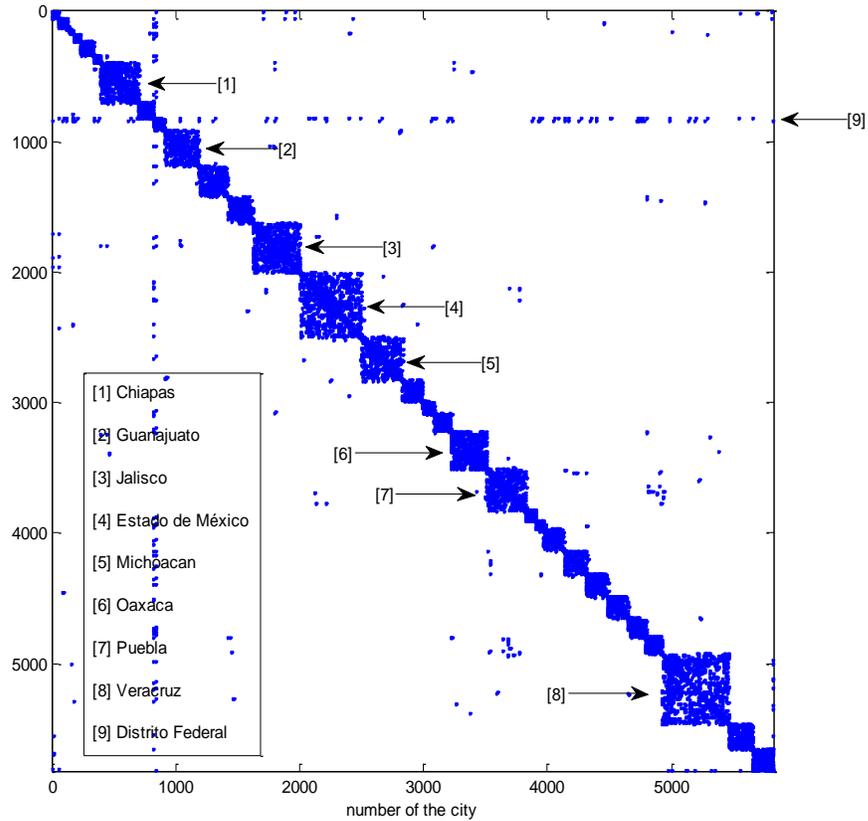

Fig. 1- A representation of the adjacency matrix. The blue dots mean coeffcients of adjacency matrix equal to one. Labels are inserted for some states of the federation. The horizontal line labeled with number 9 means the direct connections of Mexico City with other cities, which amount to 65 links.

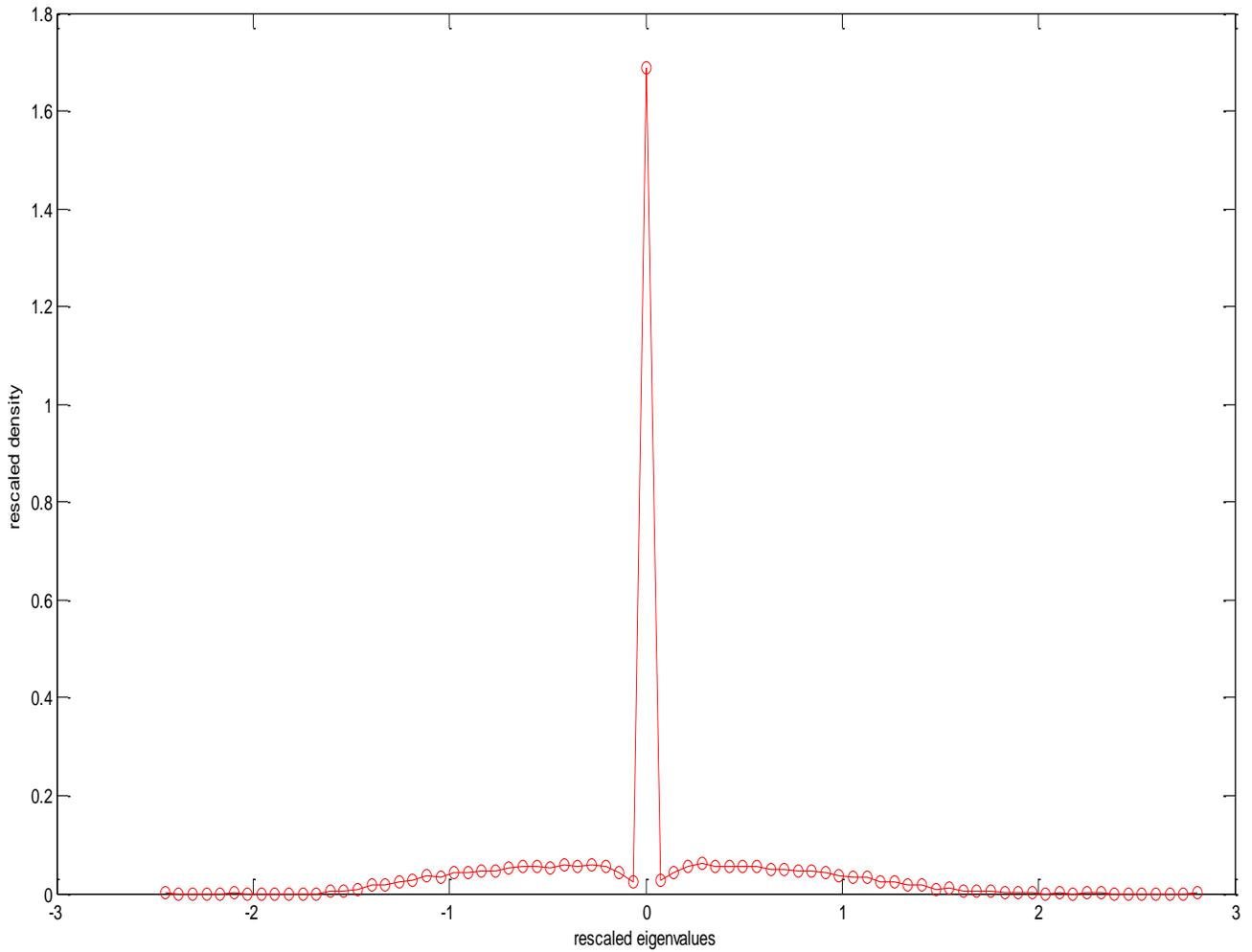

Fig. 2a- The spectral density of the adjacency matrix. The horizontal axis of the figure is rescaled as $\lambda/\sqrt{Np(1-p)}$ while the vertical axis is rescaled as $\rho\sqrt{Np(1-p)}$ where the parameter $p$ is the "implied probability of connection" (see text).

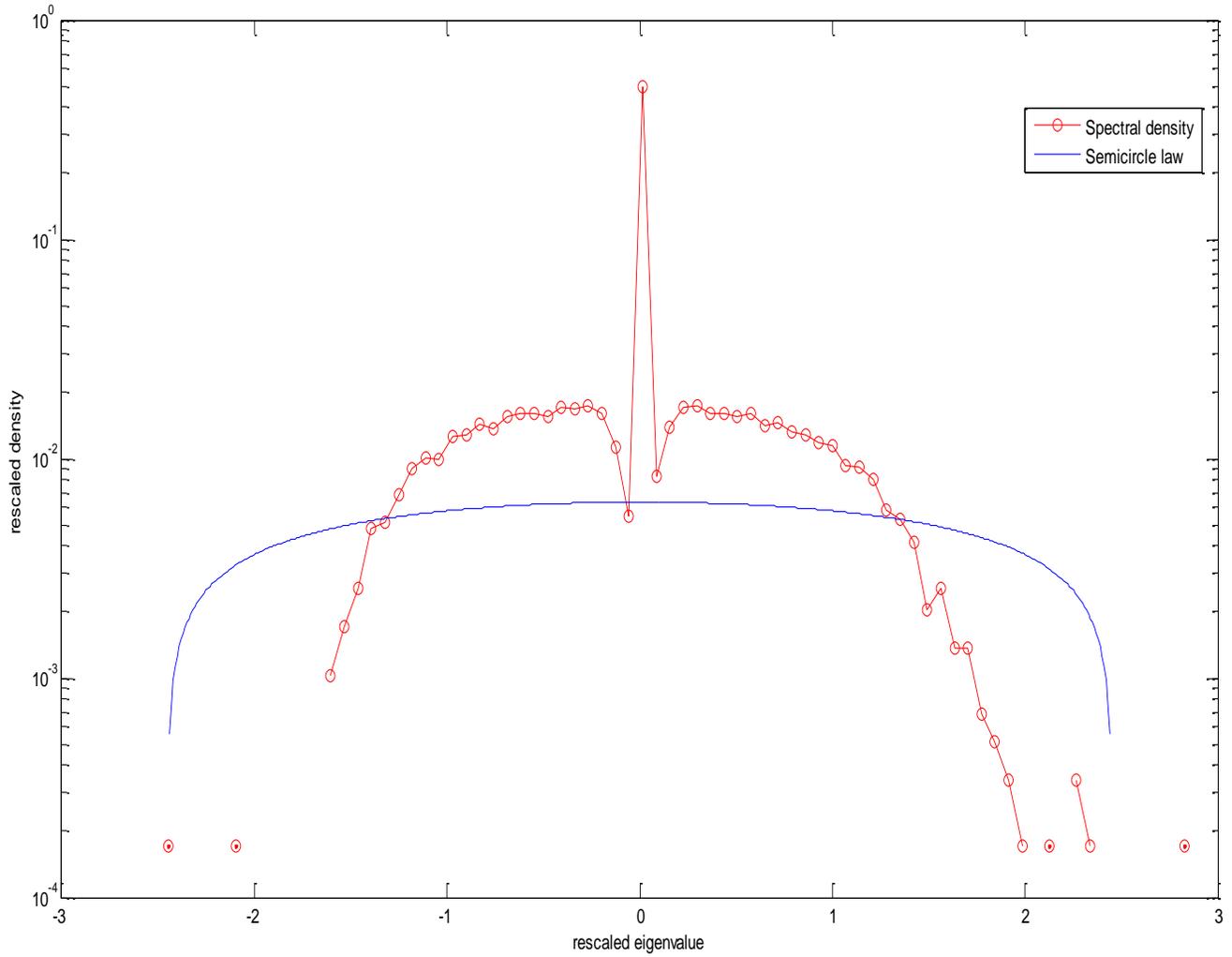

Fig. 2b- Comparison of the spectral density and semicircle law. The scale in the vertical axis is taken logarithmical in order to enhance the features of the densities.

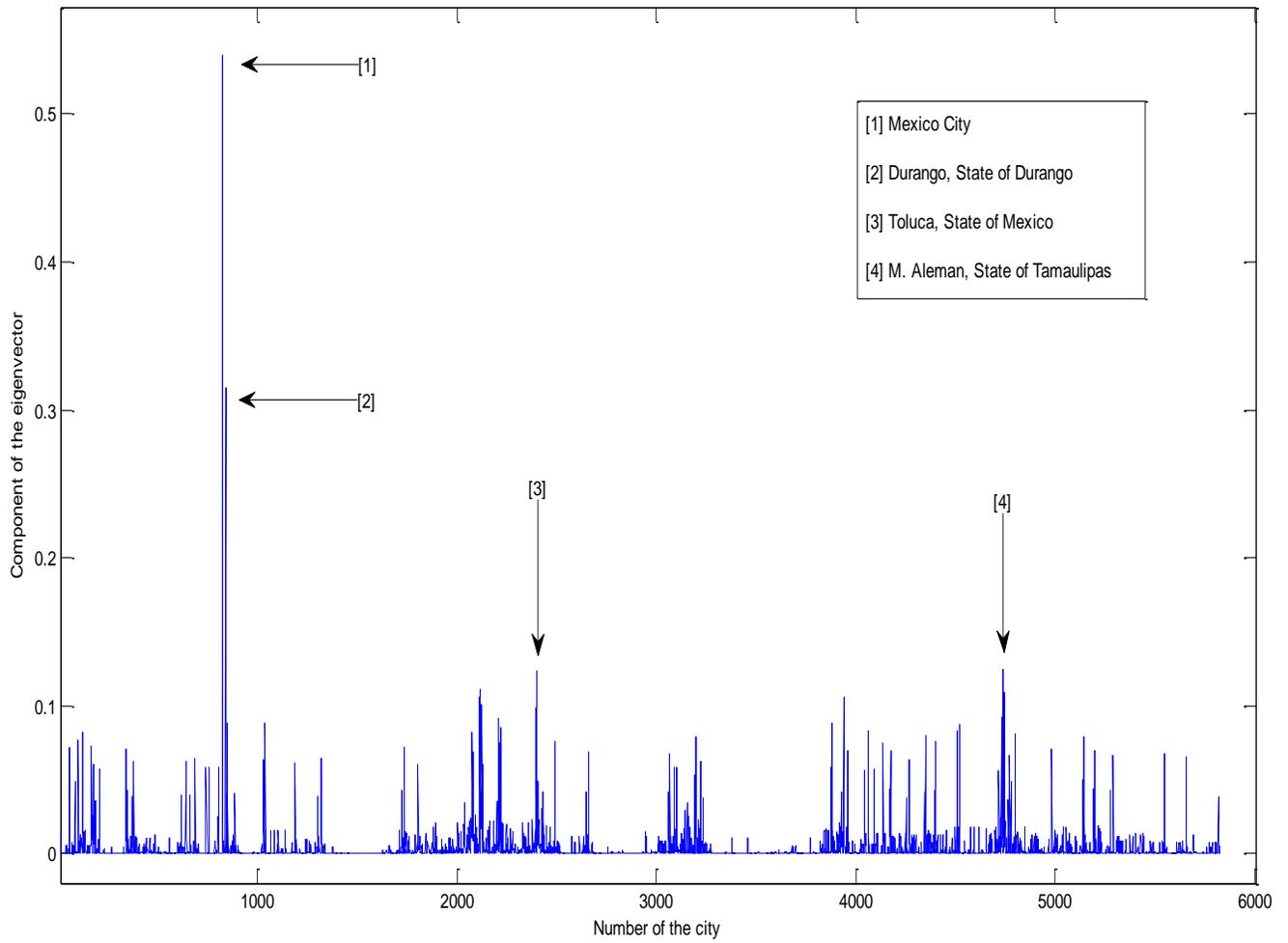

Fig. 3- Components of the eigenvector **x**. Labels are inserted for some cities. Notice the strong localization of this vector in the component corresponding to Mexico City.

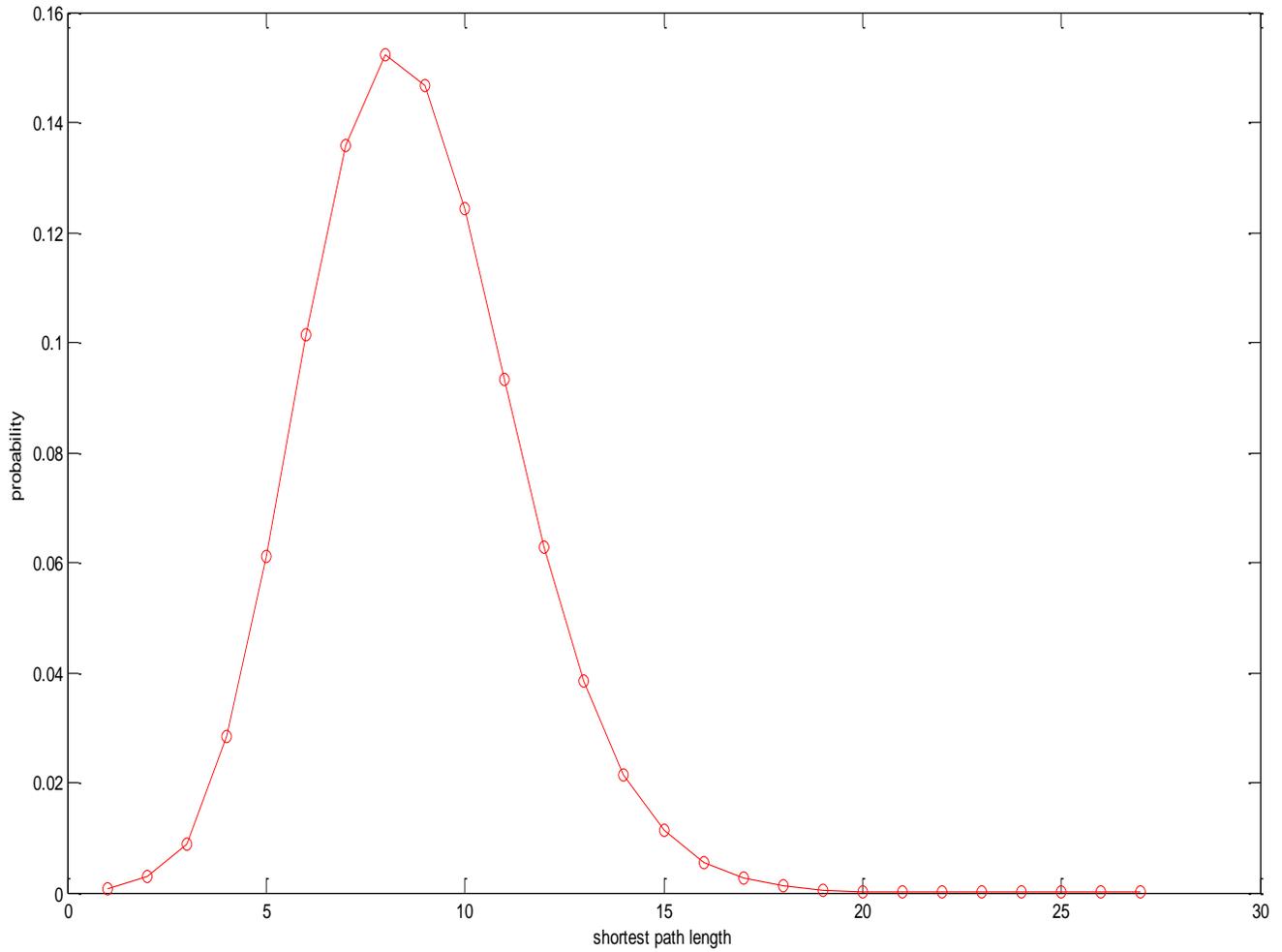

Fig. 4a- The probability density function of shortest path lengths. Notice that there is almost no skewness in the distribution.

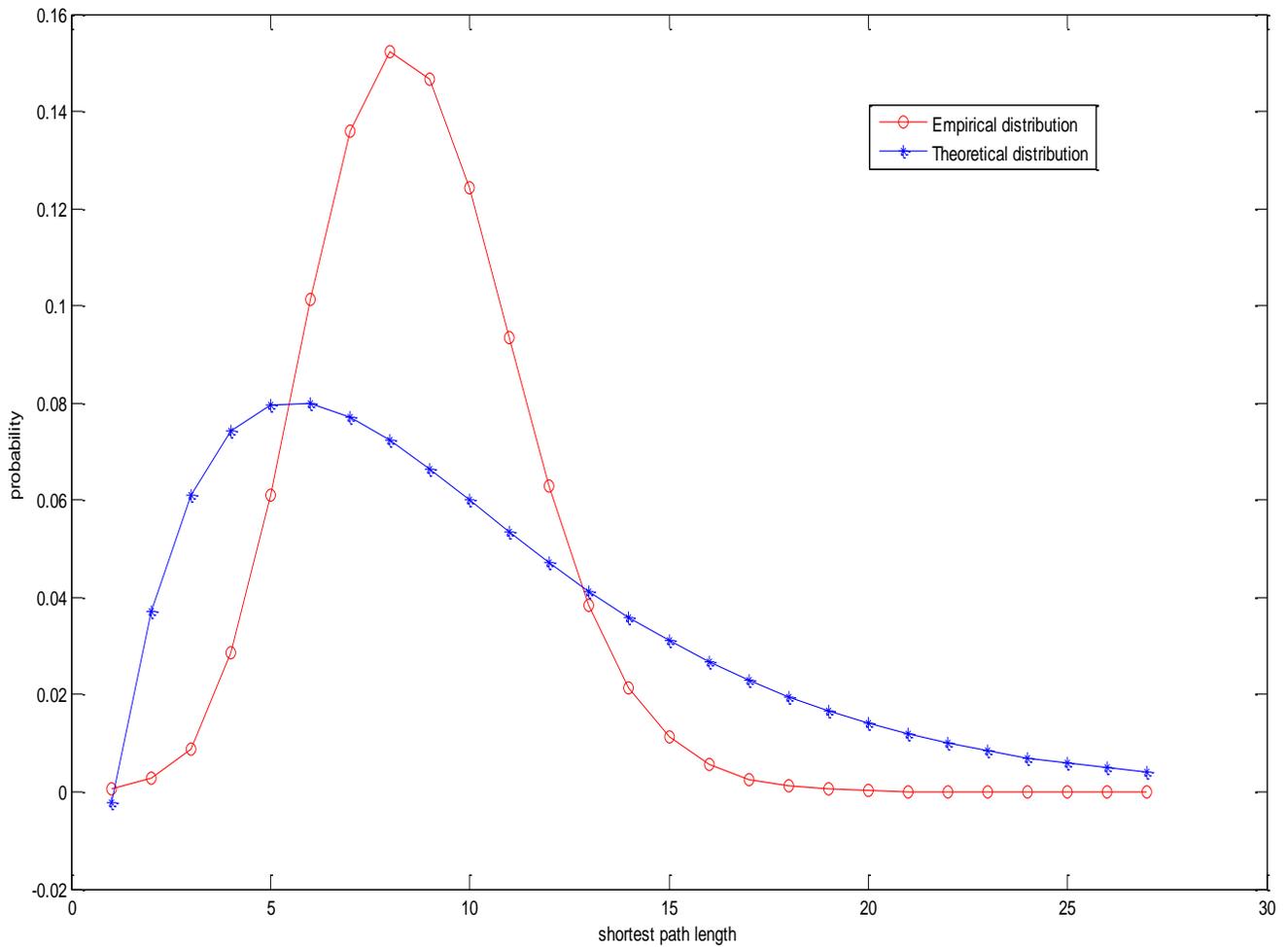

Fig. 4b- The probability density function of shortest path lengths as well as the best fit of the empirical data to the theoretical distribution (Equation (13) of [20]) proposed in that paper.

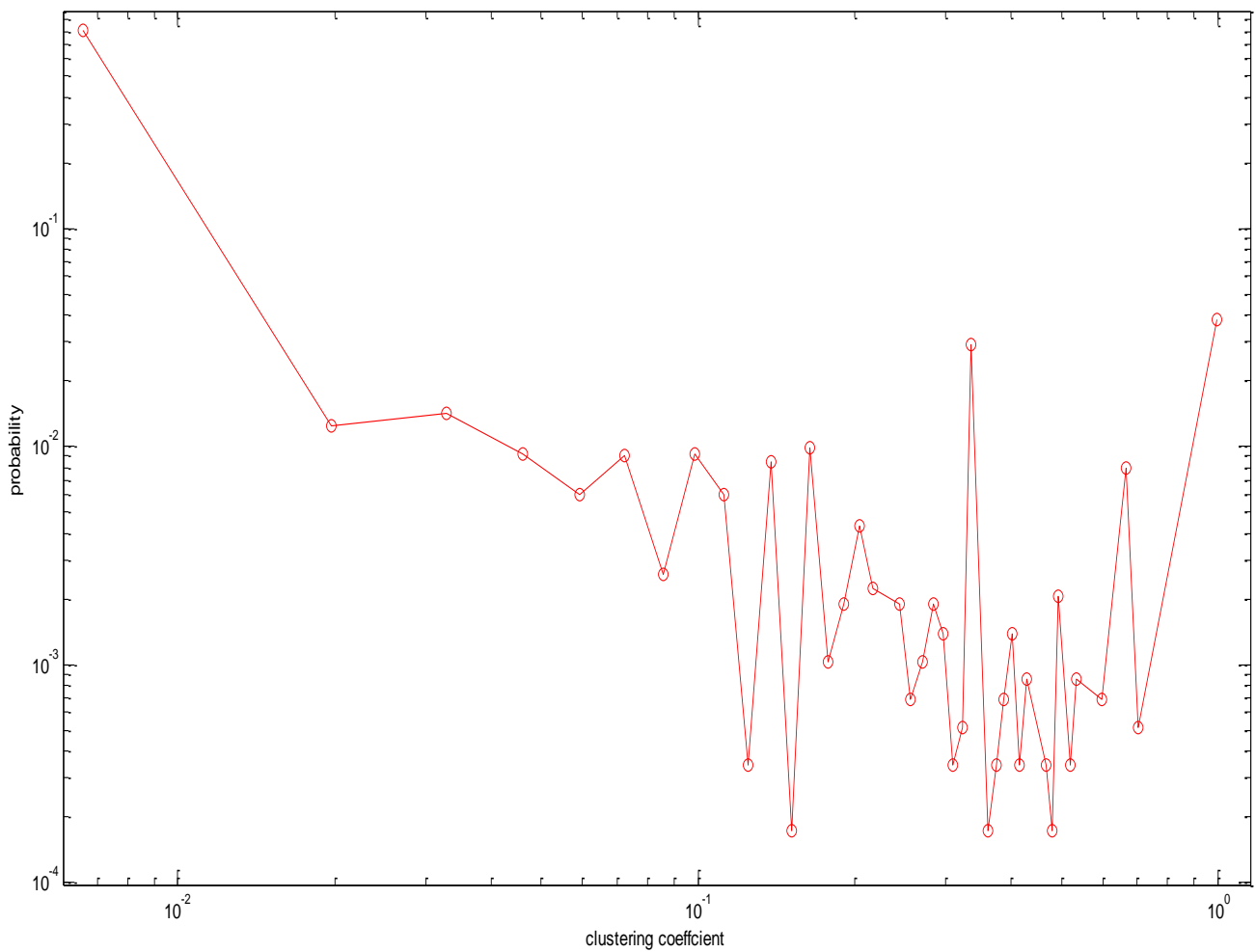

Fig. 5a- The probability density function of the clustering coefficients. Notice that about the 4% of the nodes are "cliques", i.e., all its neighbors are connected among them. The 3% of the nodes have clustering coefficient with a value about $1/3$.

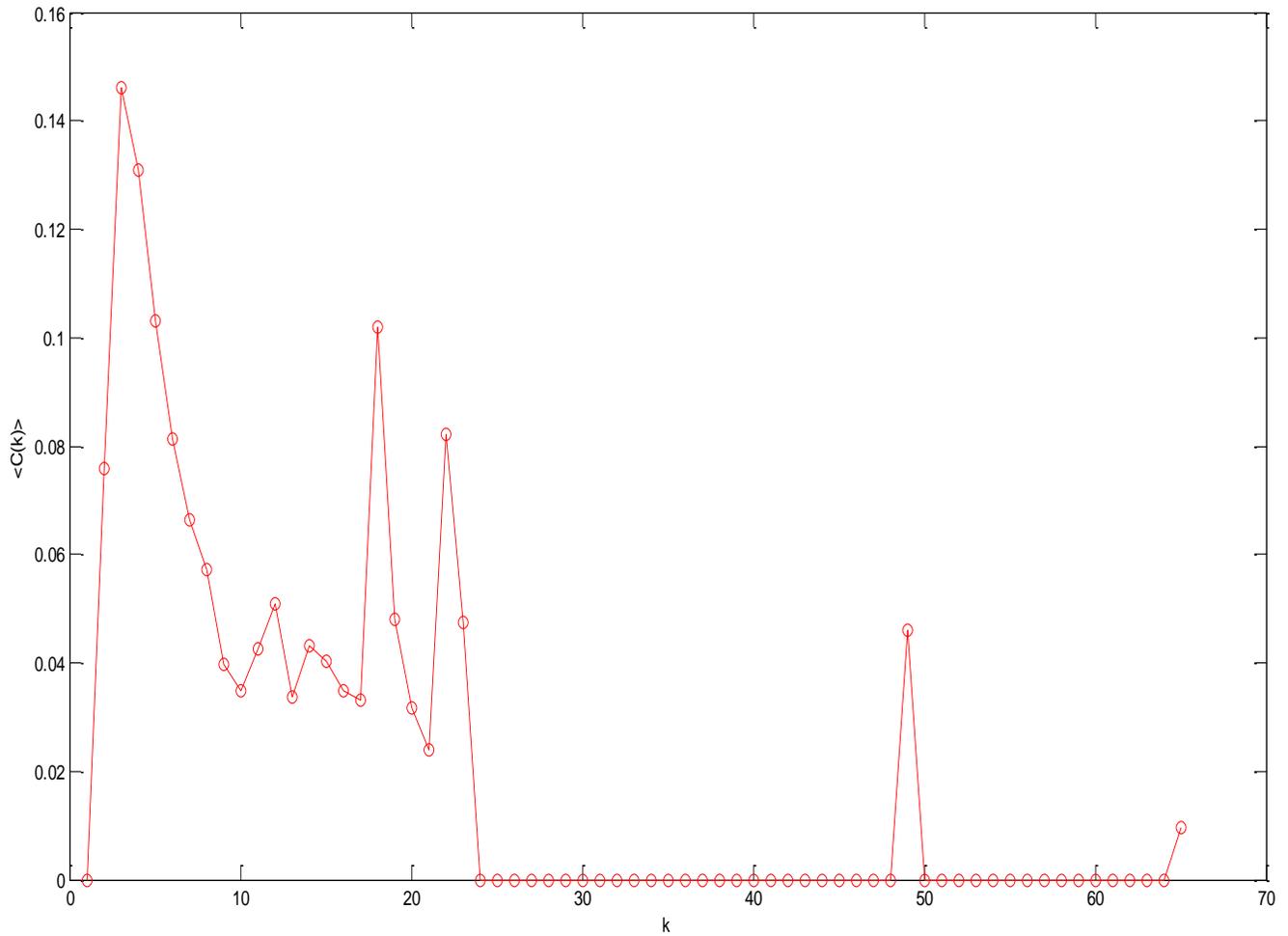

Fig. 5b- The average value of clustering coefficients with respect to the degree of the corresponding nodes. Notice that there are several values of the degree *k* which have average clustering coefficient zero. It implies a tree like behavior for these nodes.

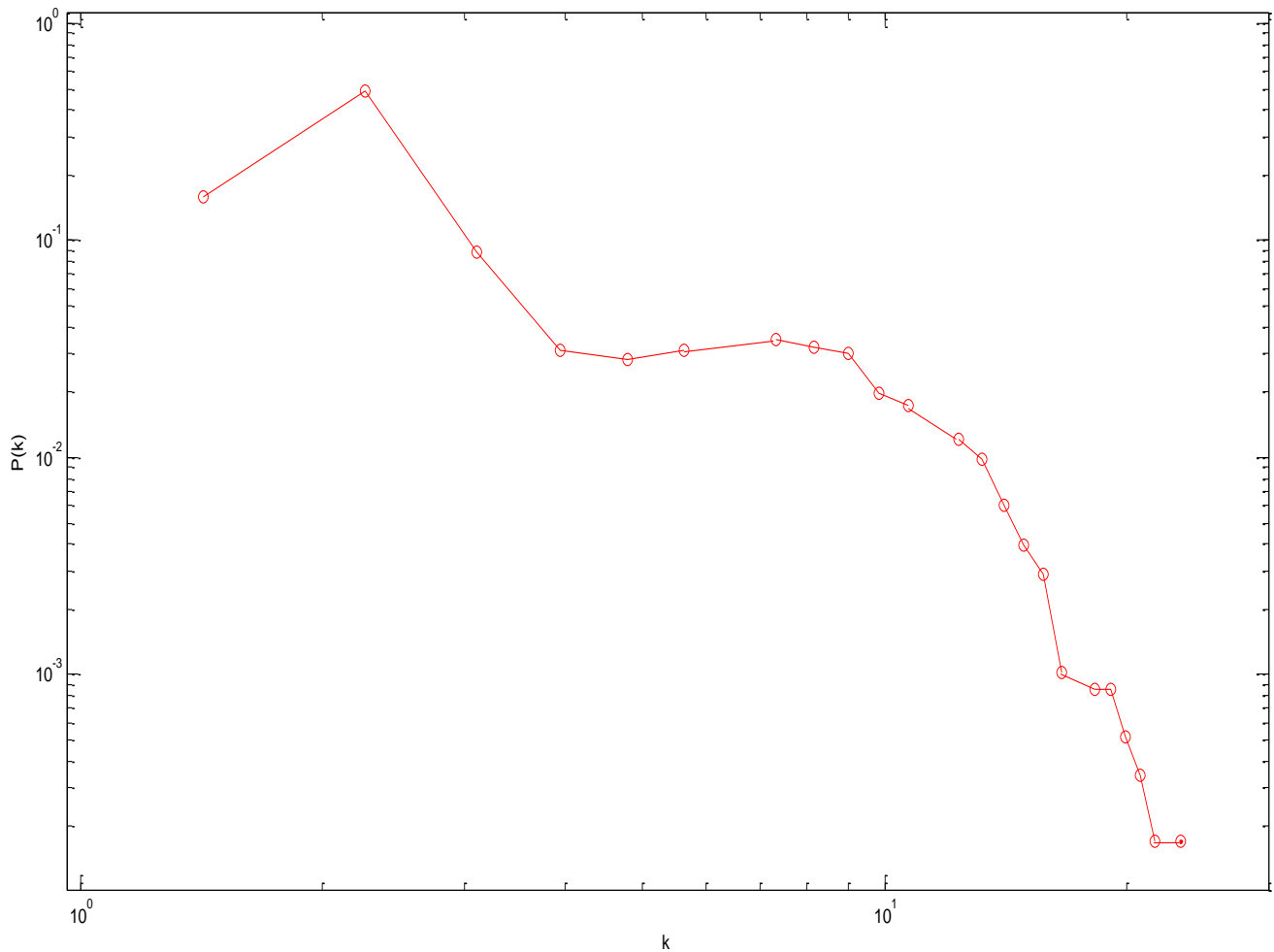

Fig. 6a- The probability density function of the node degrees. The values corresponding to Mexico City and Durango were removed because they look as outliers. No scale free or other known behavior is observed in the distribution.

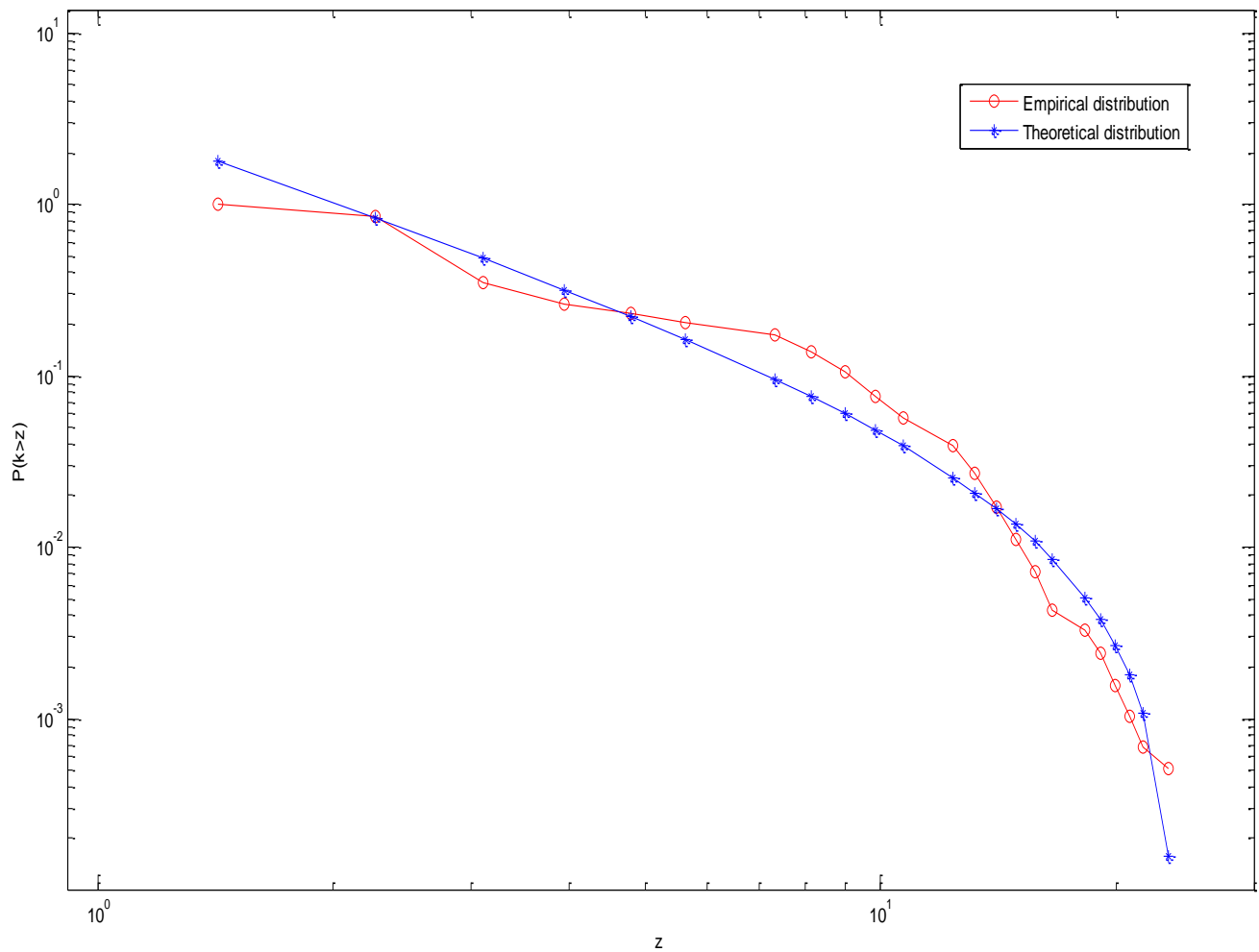

Fig. 6b- Cumulative probability distribution and theoretical distribution proposed (see text).

## Table I
## Distribution of the shortest paths

| Length of the shortest path | Number of pair of cities | Probability |
|---|---|---|
| 1 | 10312 | 0.0006193640767034 |
| 2 | 47518 | 0.0028540479244366 |
| 3 | 146139 | 0.0087774676886494 |
| 4 | 473881 | 0.0284624581101888 |
| 5 | 1017893 | 0.0611371565290746 |
| 6 | 1685940 | 0.1012617020439560 |
| 7 | 2262213 | 0.1358740754510620 |
| 8 | 2537007 | 0.1523788787960600 |
| 9 | 2443484 | 0.1467616574475800 |
| 10 | 2071338 | 0.1244096535987770 |
| 11 | 1554809 | 0.0933856517392436 |
| 12 | 1045742 | 0.0628098359484027 |
| 13 | 639825 | 0.0384294627983640 |
| 14 | 356848 | 0.0214331683517690 |
| 15 | 187519 | 0.0112628522400444 |
| 16 | 92055 | 0.0055290496587401 |
| 17 | 43260 | 0.0025983019742230 |
| 18 | 20010 | 0.0012018498036108 |
| 19 | 8534 | 0.0005125730246884 |
| 20 | 3272 | 0.0001965243656879 |
| 21 | 1227 | 0.0000736966371330 |
| 22 | 340 | 0.0000204212360434 |
| 23 | 119 | 0.0000071474326152 |
| 24 | 30 | 0.0000018018737685 |
| 25 | 17 | 0.0000010210618022 |
| 26 | 2 | 0.0000001201249179 |
| 27 | 1 | 0.0000000600624590 |